\documentclass[aps,prb,preprint,superscriptaddress,showpacs]{revtex4}

\usepackage{graphicx}
\usepackage{dcolumn}
\usepackage{bm}
\usepackage{natbib}
\usepackage{amsmath}
\usepackage{amssymb}
\usepackage{amstext}

\begin{document}

\title{Crystal to stripe reordering of sodium ions in Na$_x$CoO$_2$ ($x$ = 0.75, 0.78, and 0.92)}

\author{D.~J.~P.~Morris}
 \email{jonathan.morris@helmholtz-berlin.de}
 \affiliation{Helmholtz-Zentrum Berlin f\"{u}r Materialien und Energie, Lise-Meitner-Campus, Glienicker Str. 100, D-14109 Berlin, Germany.}
 \affiliation{Department of Physics, The University of Liverpool, Liverpool L69 7ZE, UK.}
 \author{M.~Roger}%
 \affiliation{Service de Physique de l'Etat Condens{\'e}, (CNRS/MIPPU/URA 2464), DSM/IRAMIS/SPEC, CEA Saclay, P.C. 135, F-91191 Gif Sur Yvette, France.}
\author{M.~J.~Gutmann}
 \affiliation{ISIS Facility, Rutherford Appleton Laboratory, Chilton, Didcot, Oxon OX11 0QX, UK.}
\author{J.~P.~Goff}
 \affiliation{Department of Physics, Royal Holloway, University of London, Egham, Surrey TW20 0EX, UK.}
 \author{D.~A.~Tennant}
 \affiliation{Helmholtz-Zentrum Berlin f\"{u}r Materialien und Energie, Lise-Meitner-Campus, Glienicker Str. 100, D-14109 Berlin, Germany.}
 \affiliation{Institut f{\"u}r Festk{\"o}rperphysik,
Technische Universit{\"a}t Berlin, Hardenbergstr. 36, Berlin D-10623, Germany.} 
 \author{D.~Prabhakaran}
 \affiliation{Clarendon Laboratory, Parks Road, Oxford OX1 3PU, UK.}
\author{A.~T.~Boothroyd}
 \affiliation{Clarendon Laboratory, Parks Road, Oxford OX1 3PU, UK.}
 \author{E.~Dudzik}
 \affiliation{Helmholtz-Zentrum Berlin f\"{u}r Materialien und Energie, Lise-Meitner-Campus, Glienicker Str. 100, D-14109 Berlin, Germany.}
 \author{R.~Feyerherm}
 \affiliation{Helmholtz-Zentrum Berlin f\"{u}r Materialien und Energie, Lise-Meitner-Campus, Glienicker Str. 100, D-14109 Berlin, Germany.}
\author{J.~-U.~Hoffmann}
 \affiliation{Helmholtz-Zentrum Berlin f\"{u}r Materialien und Energie, Lise-Meitner-Campus, Glienicker Str. 100, D-14109 Berlin, Germany.}
\author{K.~Kiefer}
 \affiliation{Helmholtz-Zentrum Berlin f\"{u}r Materialien und Energie, Lise-Meitner-Campus, Glienicker Str. 100, D-14109 Berlin, Germany.}


\date{\today}

\begin{abstract}
The sodium reordering in Na$_x$CoO$_2$ in the vicinity of room temperature is rationalized at high $x$ in terms of phase transitions between square and striped phases. A striking hexagon-of-hexagons diffraction pattern observed for $x$=0.78 can be reproduced using coexisting square and striped phases that are related by simple shear deformations. All compositions exhibit a partial melting transition to a disordered stripe phase just below room temperature, which alters the topology of the electrical conduction pathways.
\end{abstract}



\maketitle


Nanopatterning of sodium vacancy clusters in sodium cobaltate, Na$_{x}$CoO$_{2}$, templates the Coulomb landscape on the Co layers and, therefore, controls the electronic and magnetic properties \cite{Roger}. The concentration of sodium, $x$, can be altered electrochemically, directly changing the number of electrons in the triangular Co layers \cite{Delmas1980, Chou3}. Furthermore, for a fixed composition the physical properties depend sensitively on thermal history, since quenching and slow cool from high temperature lead to radically different heat capacity and susceptibility signals \cite{Batlogg}. These different approaches offer attractive possibilities to control the physical properties in real time.

The structure of this system creates two different environments for the sodium: Na1 which lies between adjacent cobalt ions, and Na2 which sits on a lower energy site at the center of a cobalt trigonal prism. The delicate balance between these interpenetrating hexagonal lattices causes the vacancies on the sodium layer to become attractive at short distances \cite{Roger}. Long range Coulomb interactions then allow these vacancy clusters to order over long range leading to observable superstructure peaks. This Na$^{+}$ ordering buckles the CoO$_{2}$ layer away from the occupied nearest-neighbour sodium sites (Na1) allowing cages to form within which the sodium ions are able to rattle. Meanwhile the Coulomb landscape on the cobalt layer is modified allowing narrow conduction pathways, and leading to strong correlation of the electrons. Both of these properties mean this is a phonon-glass-electron-crystal in agreement with studies of Na$_{1.2-x}$Ca$_{x}$Co$_{2}$O$_{4}$ and Li$_{0.48}$Na$_{0.35}$CoO$_{2}$ \cite{Ren, Takahata} but with a `rattler' site and this helps to explain the high thermopower in this system. Variations in the Coulomb landscape are expected to lead to the mixed valancy states as observed by NMR \cite{Mukhamedshin1, Alloul, Julien} and possible trapping of spins in the minima. Further observation and understanding of superstructures at different concentrations, and of the relevant transitions, will lead to greater insight into the overall control exerted on the system by the ionic ordering.

Whereas Ref. \onlinecite{Roger} was mainly focused on low temperature ordering, in the range $0.75<x<0.92$, we concentrate here on the evolution of the sodium order as the temperature is increased. Unexpectedly we do not observe, in the same composition range, a simple disappearance of long-range order at some critical temperature but a rearrangement of the sodiums from ``square" long range order to striped order at T=285K (as reported in \cite{Roger}, the low temperature cell is a slightly deformed square with internal angles close to but not equal to 90$^{\circ}$ that we denote by ``square cell" for simplicity). Striped order continues to exist, well above room temperature, at least up to T=350K. The reordering of the sodiums is accompanied by a change of the potential exerted by the sodium superstructure on the cobalt sites. This has a radical influence on the topology of the conducting paths -- which have now a unidimensional profile -- for the mobile carriers in the Co planes.

Single crystals of sodium cobaltate were grown using the floating zone technique \cite{Prabhak1} and then cleaved to produce high quality samples. Neutron diffraction measurements were carried out on the SXD diffractometer at ISIS and also on the flat-cone diffractometer E2 at HZB. Complementary hard X-ray diffraction on MAGS at HZB was carried out with X-ray energy of 12.398keV giving penetration of the sample allowing the bulk ordering to be probed. SXD uses the neutron time-of-flight Laue method along with 2$\pi$ steradians coverage of solid angle by position sensitive detectors to sample large volumes of reciprocal space \cite{SXDpaper}. This provides an overview of the superstructure peaks across many Brillouin zones which were then resolved at higher \textit{\textbf{Q}}-resolution using E2. Sample conditions on SXD were controlled using a closed-cycle He refrigerator and by a variable temperature cryostat on E2. Three single crystals of nominal composition $x=0.75$, 0.92 and 0.78,  were investigated at different temperatures. The superstructure for $x=0.75$ and 0.92 at 150K are reported in Ref. \onlinecite{Roger}.

\begin{figure}
   \includegraphics[width=4in]{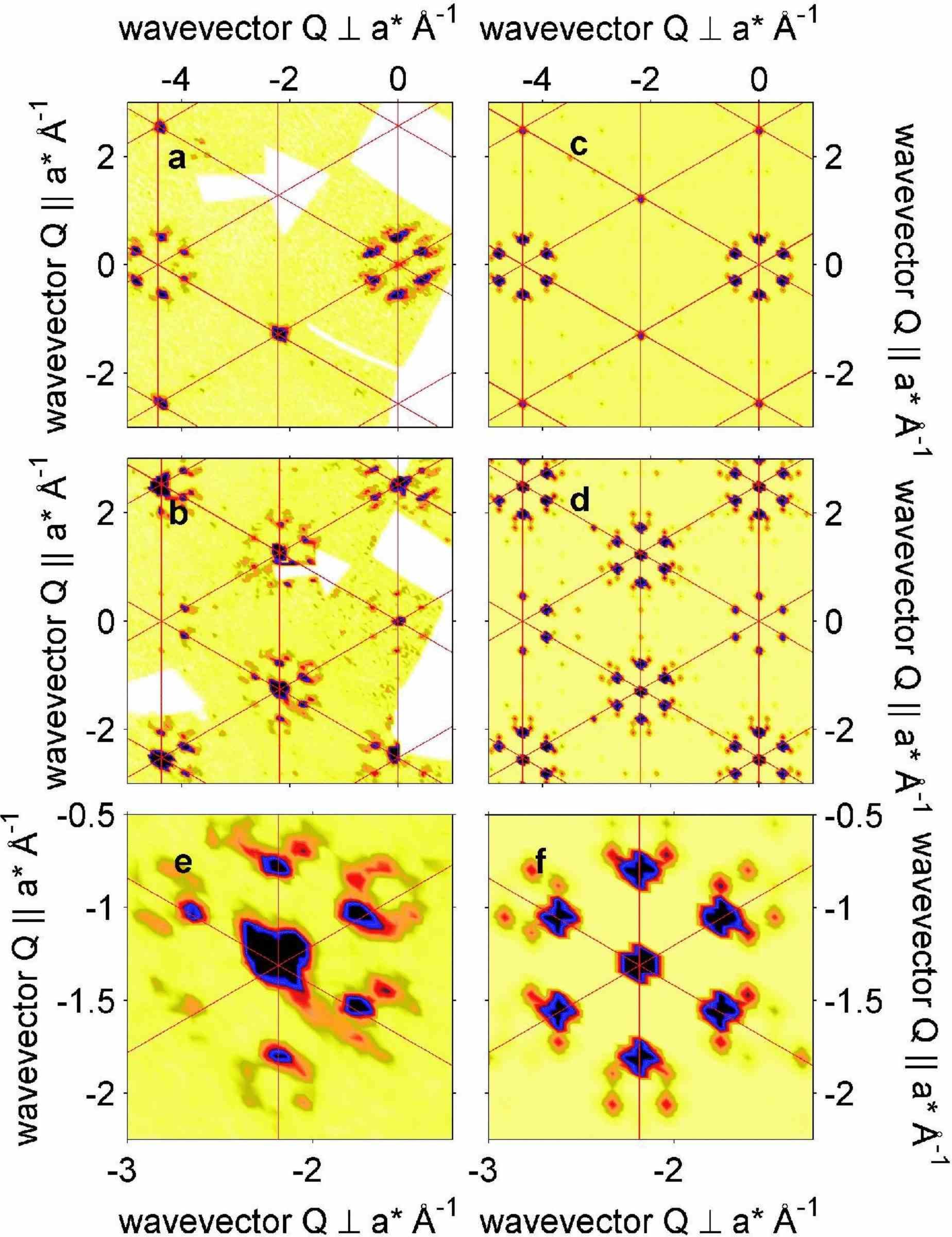}
   \includegraphics[width=2in]{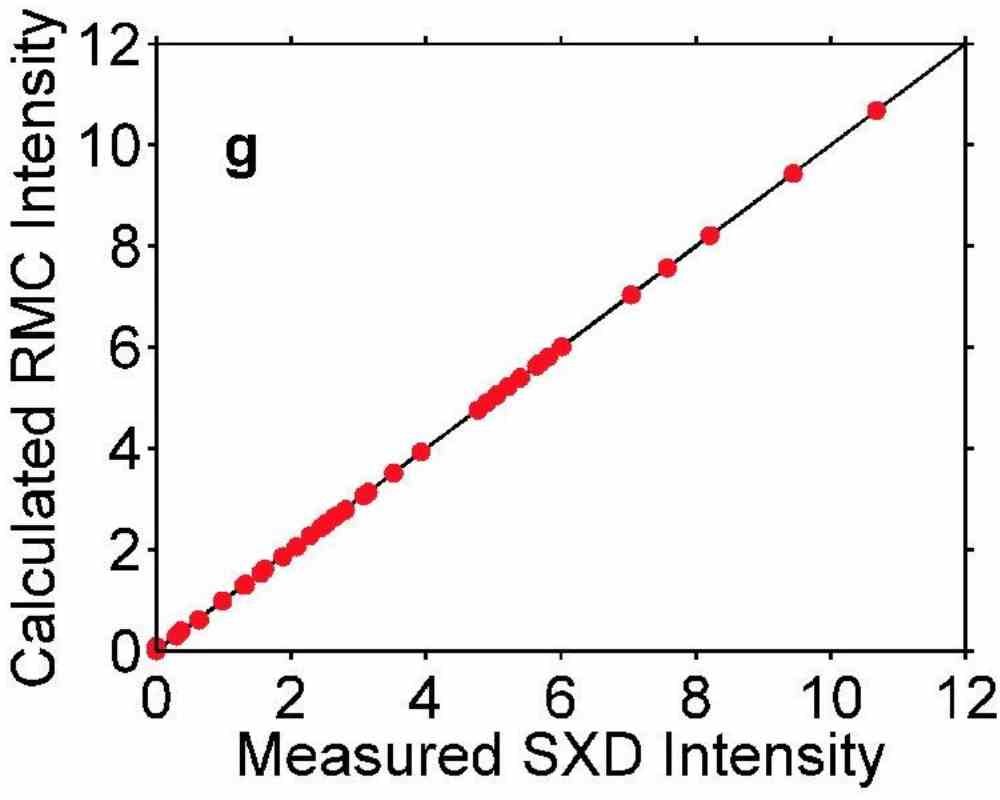} 
       \caption{(Color) (a, b) Laue diffraction data from SXD for Na$_{0.78}$CoO$_{2}$ for the $l$=11 and 7 planes respectively showing the hexagon-of-hexagons around the Bragg peak positions along with the corresponding calculations (c, d). The calculation uses phase coexistence of square and stripe phases. Detail of the agreement between data for $l$=7 (e) and calculation (f). Measured versus calculated RMC intensity for the `square' cell showing the quality of agreement, black line shows perfect fit (g).}
   \label{fig:DataCalcHex}
\end{figure}

Samples of $x=0.75$, 0.78 and 0.92 show very similar spectra at T=350K. However, at low temperature, the $x=0.78$ single crystal showed a striking ``hexagon-of-hexagons'' superstructure (Fig. \ref{fig:DataCalcHex}), different from that obtained in both $x$=0.75 and $x$=0.92 samples \cite{Roger}. Fig. ~\ref{fig:DataCalcHex}(e) focuses on the hexagon-of-hexagons superstructure pattern that surrounds the hexagonal Bragg peak positions on the $l$=7 plane for the $x$=0.78 SXD data. The data show a delicate $l$-dependence (Fig. \ref{fig:DataCalcHex}(a, b)) with the superstructure appearing around the Bragg positions that agree with $h+2k\neq 3n$ \& $l$=even over the range of $l$ measured and then around all Bragg positions for odd $l$ values below $l$=11 (superstructure peak intensities are sometimes weak around the $h+2k\neq 3n$ positions) and for $h+2k=3n$ \& $l$=11. All planes show in-plane modulations of peak intensity towards or away from the (00$l$) position indicative of buckling of the CoO$_{2}$ plane \cite{Roger}.

The superstructure peaks in Fig. \ref{fig:DataCalcHex} for $x$=0.78 at T=150K can be indexed on a ($\frac{1}{15}\mathbf{a^{*}}\times\frac{1}{15}\mathbf{b^{*}}$) grid, and the integrated intensities have been analysed by Reverse Monte Carlo (RMC) using simulated annealing in the Canonical Ensemble. The (fictitious) energy $E(X)$ is defined in terms of the mean-square deviations of the experimental intensities from the calculated structure factors for a given 3D configuration $X$ of all particles in the unit cell. The Metropolis algorithm is used with hopping of a sodium ion from an occupied to a vacant site with small displacements of surrounding ions. The temperature is slowly decreased, and the configuration at T=0 corresponds to the least-square fit of the integrated peaks. 

\begin{figure}
   \includegraphics[width=2in]{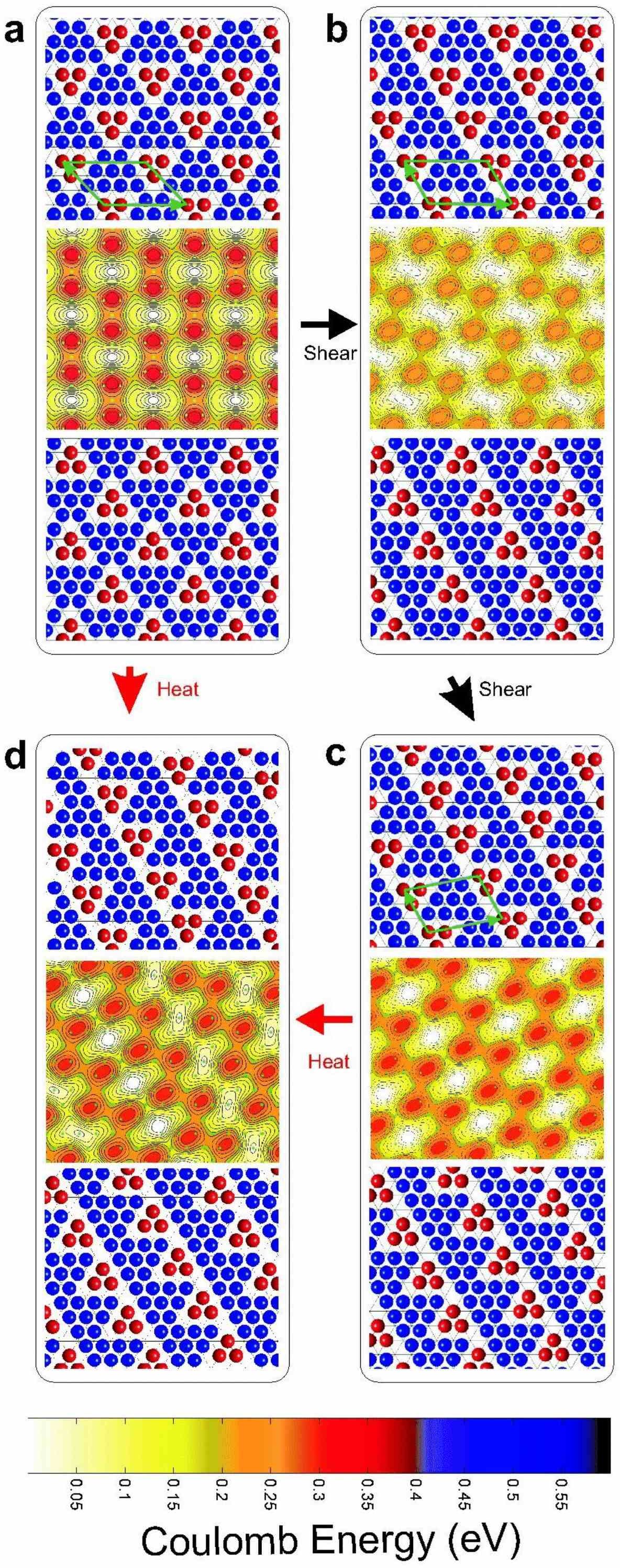}
       \caption{(Color) Each panel consists of the ionic ordering on adjacent sodium layers (red spheres show occupied Na1 sites and blue spheres shows Na2 sites as described in the text), in phases reported here, separated by the Coulomb landscape on the intervening cobalt layer (white/yellow show conduction pathways). a) The square trivacancy cluster structure seen in Na$_{0.75}$CoO$_{2}$ \cite{Roger} requires a shear distortion parallel to the $a$-lattice direction to take it to (b) an unobserved stripe lattice and then a further shear along $b$ to modify into (c) the stripe structure observed in Na$_{0.78}$CoO$_{2}$. (d) All three concentrations show the same disordered stripe structure above the melting transition at 285K.}
   \label{fig:Shear}
\end{figure}

A fraction of the peaks can be identified as coming from the square lattice described in Ref. \onlinecite{Roger} and shown in Fig. \ref{fig:Shear}(a), and these were excluded from the RMC calculations. A very robust result from RMC simulations is that the absence of significant intensity for most superlattice reflections away from the hexagonal reciprocal lattice points, leads naturally to multi-vacancy cluster patterns. For $x$=0.78 at T=150K we obtain stripes of trivacancies condensing into long-range order with cell vectors $\mathbf{a'}=5\mathbf{b}-\mathbf{a}$ and $\mathbf{b'}=5\mathbf{b}-4\mathbf{a}$, as shown in Fig. \ref{fig:Shear}(c) with vacancy clusters on the adjacent Na layers being as far apart as possible to reduce Coulomb energy giving 3D order. This in-plane stripe structure is a simple modification of the square lattice requiring only two shear distortions, of one lattice spacing each, to the supercell to reach this end structure (figure~\ref{fig:Shear}).The buckling of  the CoO$_{2}$ plane induced by the sodium superstructure has a maximum distortion along the c axis of 0.025$c$; the effect of this buckling on the screening of the Coulomb potential is negligible. Fourier transforms of this striped structure are then added to the Fourier transforms of the square cell in a ratio of $\frac{I_{Stripe}}{I_{Square}}$=1.3. This summation produces remarkable agreement with this complex data set (see comparison in fig \ref{fig:DataCalcHex}). Full details of the RMC simulations will be reported elsewhere. A number of experiments \cite{Geck, Chou2008} provide evidence for the presence of vacancy clusters, and the formation of these clusters is understood in terms of topological constraints and Coulomb repulsion \cite{Roger, Meng2}. This is a first-order effect involving energies of a fraction of one eV. Long-range ordering of these clusters is a more subtle effect. Refs. \onlinecite{Roger} and \onlinecite{Meng2} agree on the possible occurrence for 0.75$<x<$0.85 of a number of long-range commensurate superstructures with ground-state energies differing by only a few meV, which is not significant with respect to second-order effects such as cooperative phenomena involving coupling to phonons and mobile carriers in Co layers.  We determine unambiguously which, among those possible superstructures, the real system chooses. The full theoretical interpretation of the results, especially our finite temperature transition, is a highly challenging problem.

\begin{figure}
   \includegraphics[width=3in]{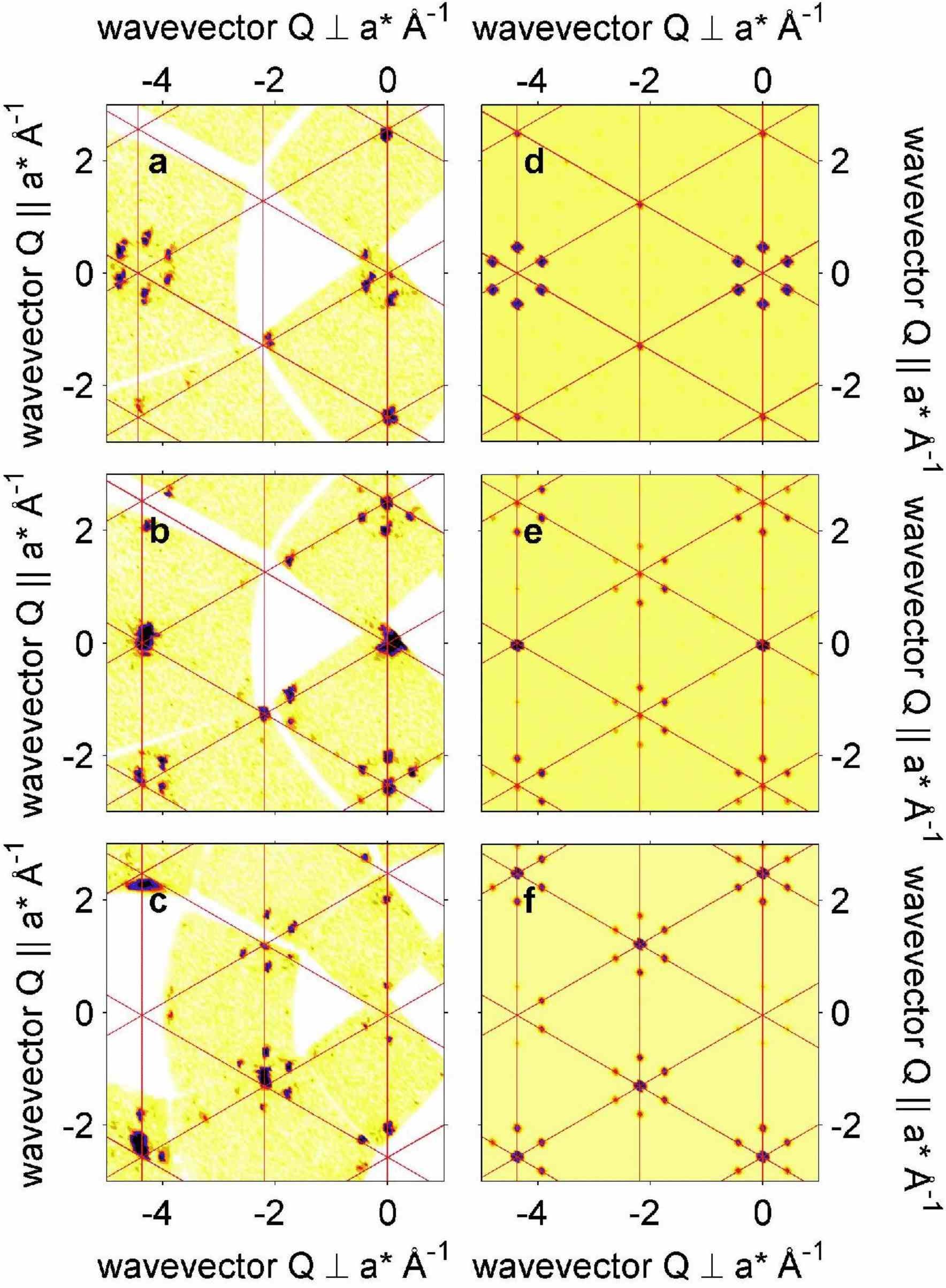}
       \caption{(Color) Above 285K, in all three concentrations ($x$=0.75, 0.78 and 0.92), the long range ordered structure (figs. \ref{fig:Shear}(a), (c)) melt into one with disorder between stripes giving the same superstructure pattern (fig. \ref{fig:Shear}(d)). (a) $l$=11 for Na$_{0.92}$CoO$_{2}$ at 350K. (b) $l$=10 and (c) $l$=7 showing the $l$ dependence. (d-f) Calculation using the disordered stripe structure.}
   \label{fig:DataCalc2}
\end{figure}

The patterns at 350K look quite different from those of Ref. \onlinecite{Roger}, at T=150~K, with changes of symmetry suggesting a phase transition. The superstructure pattern above this transition can be labeled on a ($\frac{1}{5}\mathbf{a^{*}}\times \frac{1}{5}\mathbf{b^{*}}$) grid. Figure \ref{fig:DataCalc2} shows the $l$=7, 10, 11 cuts obtained for the $x$=0.75 samples at  T=350K. Again the pattern shows a similar delicate $l$ and in-plane dependence.

\begin{figure}
   \includegraphics[width=4in]{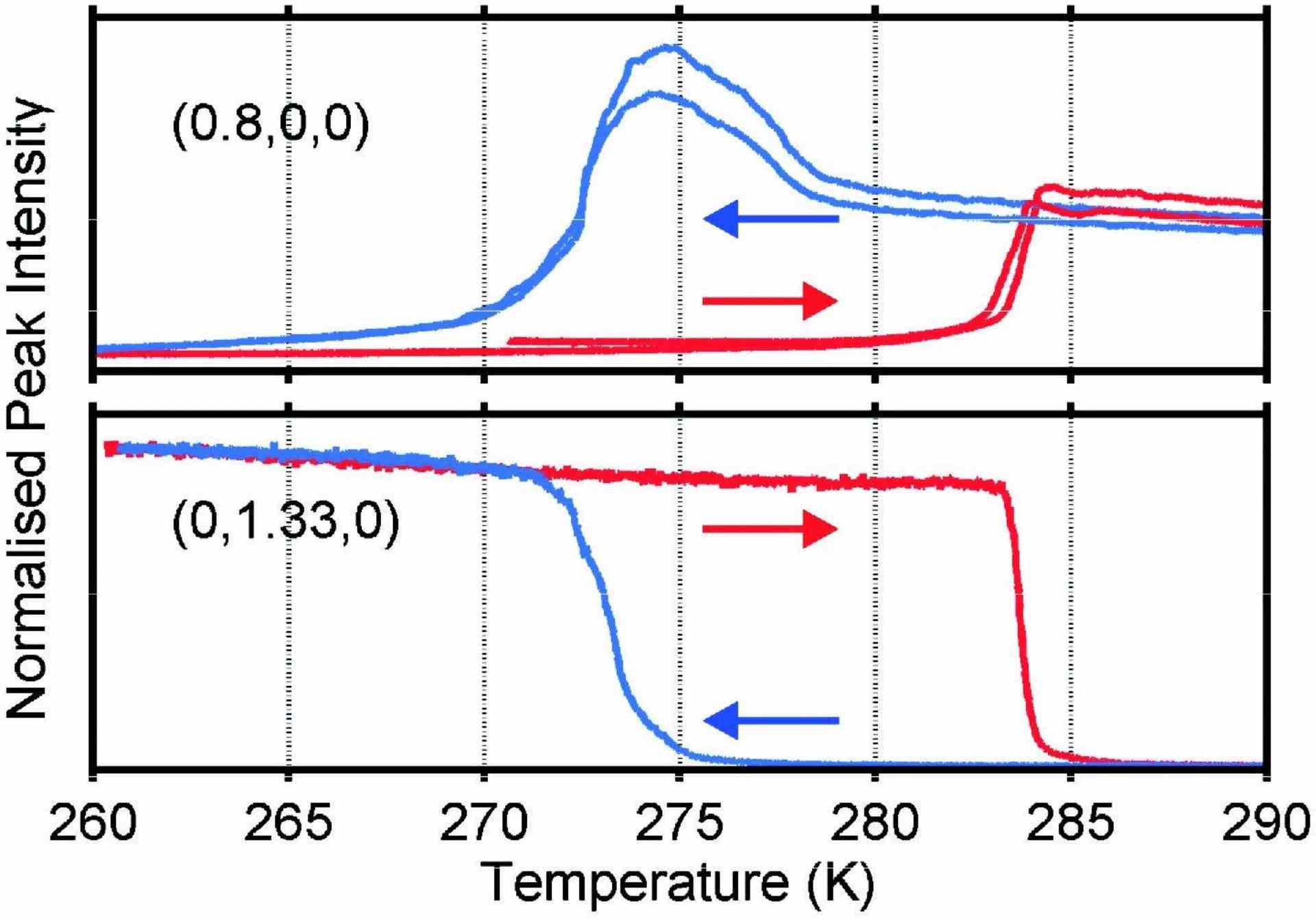}
       \caption{(Color) Hysteresis behaviour of the ordered phase to disordered stripe transition in the intensities of superstructure peaks measured using hard X-ray diffraction on $x$=0.75.}
   \label{fig:Transition}
\end{figure}

X-ray data shows a pronounced hysteresis of over 10K between heating and cooling runs (Fig. \ref{fig:Transition}). Here \textit{\textbf{Q}}=$(0, 1.33, 0)$ is from the square phase in Fig. \ref{fig:Shear}(a), whereas the peak at \textit{\textbf{Q}}=$(0.8, 0, 0)$ is where the scattering is expected for either of the striped phases in Figs. \ref{fig:Shear}(c) and (d). There is a highly unusual peak in the cooling run for \textit{\textbf{Q}}=$(0.8, 0, 0)$.

The trivacancy clusters can sit on three possible sites within the stripes in Fig. \ref{fig:Shear}(c). Consider a model where the ordering of stripes is long range, but the ordering of trivacancy clusters within the stripes is not coherent from one stripe to the next. The translation vectors for this phase are $\mathbf{a'}=5\mathbf{b}+(\eta -1)\mathbf{a}$ and $\mathbf{b'}=5\mathbf{b}+(\eta -4)\mathbf{a}$ where $\eta=0,\pm 1$ at random. A section of this partially disordered structure is shown in Fig. \ref{fig:Shear}(d). Simulations of the scattering intensity from this phase are in excellent agreement with the experimental data for $x$=0.75, 0.78 and 0.92 at elevated temperature, see for example Fig. \ref{fig:DataCalc2}. Therefore, the phase transformation is a melting transition from the long-range ordered square or striped phases in Figs. \ref{fig:Shear}(a) and (c) respectively, to the disordered stripe phase in Fig. \ref{fig:Shear}(d). The anomalous peak in the intensity of the reflection at \textit{\textbf{Q}}= $(0.8, 0, 0)$ in the cooling run below 278K in Fig. \ref{fig:Transition}(a) can also be explained within this framework. If the disordered stripes in Fig. \ref{fig:Shear}(d) transform to the intermediate ordered stripe phase in Fig. \ref{fig:Shear}(c), stronger scattering is expected at the same position in reciprocal space, before the reordering to the square phase in Fig. \ref{fig:Shear}(a) below 274K.

We note that a different model with quadrivacancy clusters was previously proposed for the high-temperature phase \cite{Roger}. However, the model proposed here is a more satisfactory explanation due to the simplicity of the phase transformations in fig. \ref{fig:Shear} and also agrees with quantitative RMC analysis. All phases have the same multi-vacancy cluster type (trivacancy) and the concentration $x$ in each phase is the same.

Finally, the entropy associated with the random positions of stripes with respect to one another is $\Delta S\approx k_{B} N_{stripes} \ln 3$, where $N_{stripes}\approx\sqrt{N}$  represents the number of stripes ($N$ is the number of sites in a Na plane) and there are three different positions within stripes. Hence the entropy per particle $\frac{\Delta S}{N}\approx \frac{N_{stripes}}{N} \approx \frac{1}{\sqrt{N}}$ vanishes in the thermodynamic limit. Here the disorder is static, but some additional entropy could arise from other static (due to defects) or dynamic disorder. Additional disordering of the stripes and longitudinal disordering of the clusters within stripes may come into play forming something analogous to a smectic phase. The relatively small change in entropy involved in the transition, and any associated electronic transitions, are indicated by the lack of any large anomaly accompanying the transition.

This structural transition is accompanied with a topological change of the Coulomb landscape from 2D to 1D. Above the transition, there are lines of Co sites at the top of the wells (red stripes in Fig. \ref{fig:Shear}d) which are expected to remain Co$^{3+}$ while the conducting holes will move along the bottom of the wells (white/yellow stripes). Since the corresponding depth, of order 200meV is much larger than the hopping energy $t=$10meV, we expect at room temperature ($k_{B}T\approx$26 meV) a strong confinement of the carriers along the yellow stripes. Patterning of the Coulomb energy could also describe the observation of surface superstructure by Pai \textit{et al.} in which trimers of Na are seen to order in regular arrays \cite{Pai}. This can be thought as Na sitting on top of the energetically favourable vacancy sites in the underlying bulk ordering. Neutron diffraction is not sensitive enough to observe magnetic superstructure peaks since this will have an intensity of the order of a hundredth that of the magnetic Bragg peak from ref. \onlinecite{BayrakciMag}. The structural transition that we observed through neutron diffraction could explain some of the anomalies seen within the 270-290K temperature range: e.g. broad bump in susceptibility at 285K in $x$=0.82 \cite{Bayrakci}, transition in optical ellipsometry at 280K in $x$=0.82 \cite{Bernhard}, phonon lifetime in infrared conductivity below 295K over a range of concentrations \cite{Lupi} and anomaly in resistivity \cite{Wooldridge1, Ikeda, Roger}.

In conclusion, using neutron Laue diffraction and hard X-ray diffraction, we have observed square to stripe reordering of the sodium ions near room temperature. As a consequence the mobile carriers have to follow restricted paths in the Co planes with a topological change from 2D to 1D at the transition. This is expected to lead to changes in the topology of the Fermi surface, suggesting ARPES measurements to test this prediction. This 1D character of the conducting path offers an alternative direction in developing new materials with improved thermoelectric properties. The possibility of producing single domains using electric fields or by thin-film deposition should be investigated since the electrical conductivity will be higher along the stripes than for multidomain samples, increasing the figure of merit for thermoelectric applications. The idea to improve the figure of merit by reducing the dimensionality has been suggested a long time ago \cite{Hicks}. Here we find a self-organising system with 1D conducting stripes on the nanoscale.

\bibliographystyle{apsrev}

\end{document}